\documentclass[11pt]{article}
\usepackage{epsfig}
\def\question#1 {~\\ {\bf\it #1 }\\}
\def\question#1 {}
\newcommand{\A}{{\mathcal A}}

\newcommand{\be}{\begin{equation}}
\newcommand{\ee}{\end{equation}}
\newcommand{\ba}{\begin{eqnarray}}
\newcommand{\ea}{\end{eqnarray}}
\newcommand{\bb}{\begin{thebibliography}}
\newcommand{\eb}{\end{thebibliography}}
\newcommand{\ci}[1]{~\cite{#1}}
\newcommand{\lab}[1]{\label{#1}}

\newcommand{\ahnf}{${\mathcal A}^h_{nf}$}

\begin{document}

\title{
 Large-distance effects on spin observables at RHIC\footnote{presented by O.V.S. at the 16th International Spin Physics Symposium, spin2004, 
       October 10-16 2004, Trieste, Italy}  }

\author{O.V. Selyugin\footnote{Inst. de Physique,
Univ. de Li\`ege,
Li\`ege, Belgium}~\footnote{BLTPh,
 JINR, Dubna,
 Russia.}~, J.-R. Cudell$^\dagger$ and E. Predazzi\footnote{Dip. di Fisica Teorica - Univ. di Torino
   and Sezione INFN di Torino, Italy.}}

\maketitle

\begin{quote}
{\bf Abstract}
{
The impact of large-distance contributions on the behaviour
of the slopes of the spin-non-flip and of the spin-flip amplitudes
is analysed.
It is shown that the long tail of the hadron potential
in impact parameter space
leads to a larger value of the slope for the spin-flip
amplitude (without the kinematic factor $\sqrt{|t|}$) than for
the spin-non-flip amplitude.
This effect
is taken into account in the calculation of the analysing power
in  proton-nucleus reactions at high energies.
}
\end{quote}
~\bigskip

The recent data from RHIC and HERA indicate
that, even at high energy, the hadronic amplitude has
a significant spin-flip contribution, ${\mathcal{A}}^{h}_{sf}$,
which remains proportional to the spin-non-flip part, \ahnf,
as energy is increased.
Theoretically, when large-distance
contributions are considered, one can obtain  a more complicated
spin structure for the pomeron coupling.
The dependence of the hadron spin-flip amplitude on the momentum transfer
at small angles is tightly connected with the basic structure of
hadrons at large distances. We show that
the slope of the ``reduced'' hadron spin-flip amplitude
({\it i.e.} the hadron spin-flip amplitude without the kinematic
factor $\sqrt{|t|}$)
can be larger than the slope of the hadron spin-non-flip amplitude
as was observed long ago~\cite{predaz,wak}.
This large slope has a small effect on the differential hadron cross section
and on the real part of hadron non-flip amplitude~\cite{sel-sl}.

The helicity amplitudes can be written
$$ \Phi_i(s,t) = \phi^h_{i}(s,t)
        + \phi_{i}^{em}(t) \exp[i \alpha_{em} \varphi_{cn}(s,t)],\ i=1,...5$$
where $\phi^h_{i}(s,t)$ comes purely from strong interactions,
$\phi_{i}^{em}(t)$  from electromagnetic interactions
($\alpha_{em}=1/137$ is the electromagnetic constant)
and
$\varphi_{cn}(s,t)$ is the electromagnetic-hadron interference phase factor.
The ``reduced'' spin-flip amplitudes are defined as
  ${\tilde\A^{h}_{sf}}(s,t) =  \phi^{h}_{5}(s,t)/(s \sqrt{|t|})$  and
  ${\tilde\A^{c}_{sf}}(s,t) =  \phi^{em}_{5}(s,t)/(s \sqrt{|t|})$.
  As usual, we define the slope $B$ of the scattering amplitudes as
   the derivative of the logarithm of the amplitudes with respect to $t$.

The contribution of large distances was studied in~\cite{tur1,cps1}.
We present here the results of a
a numerical calculation of the relative contributions of small and large
distances.  We calculate the scattering amplitude in the Born
approximation for form factors exponential or Gaussian in impact parameter
space, as a function of the upper limit $b$ of the corresponding integral
\ba
\phi^{h}_{1}(t) &\sim&  \int_{0}^{b} \ \rho \ d\rho
 \ J_{0}(\rho \Delta) f_{n} , \ \
\phi^{h}_{5}(t)/q \sim  \int_{0}^{b} \ \rho^2 \ d\rho
 \ J_{1}(\rho \Delta) \  f_{n}  . \lab{f5ab}
\ea
with $f_{n}= \exp{[-(\rho/5)^n]}$, and $n=1,2$.
We then  calculate the ratio of the slopes of these two amplitudes
$R_{BB} =B^{sf}/B^{nf}$ as a function of $b$ for these two values of $n$.
The result is shown in Fig.1.
We  see that at small $b$ the value of $R_{BB}$ is
practically the same in both cases.
However, at large distances, the behaviour of $R_{BB}$ depends on the form
factor: in the Gaussian case, $R_{BB}$ reaches its asymptotic value
($=1$) quickly, but in the exponential case,
it reaches its limit $R_{BB}=1.7$ only at large distances.
These calculations confirm our analytical analysis of the asymptotic
behaviour of these integrals at large distances.
\vglue 1.cm
\begin{figure}
\epsfysize=2.5cm
\centerline{\epsfbox{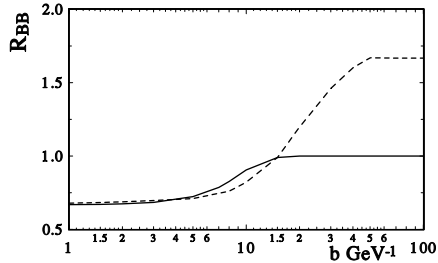}}
\caption{ The ratio $R_{BB}$ of the effectively slopes for $n=1$
(dashed line) and for $n=2$ (solid line)  the upper bound $b$ of the integrals.}
\end{figure}

We can now use these results in the description of the analysing power at
small momentum transfer, for which
there are very few high-energy data.
We take the hadron spin-non-flip amplitude
in a simple exponential form, normalised to the total cross
section~\cite{prd-comp}, and
with a size and energy dependence of the slope proportional to
their  values from $pp$-scattering~\cite{selex}.

We assume that the slope slowly rises with $\ln{s}$ in a way
similar to the $pp$ case, and normalise it
so that the spin-non-flip  amplitude  has a slope of
$62$~GeV$^{-2}$ at $p_L= 600$~GeV/c and $|t|=0.02$~GeV$^2$.
We parametrise the spin-flip part of $p ^{12}C$ scattering as
\ba
{\A^{h}_{sf}}(s,t)&=&   (k_2 \ + \ i \  k_1)
  { \sqrt{|t|} \ \sigma_{tot}^{pA}(s)\over 4\pi}
\exp\left({ B^{-}\over 2}t\right)
\ea
 According the above analysis, we investigate two extreme cases for the
 slope of the spin-flip amplitude: I -
the spin-flip and the spin-non-flip amplitude have the same slope
 $B^{-} \ = \ B^{+}$;  II - $B^{-} \ = \ 2 B^{+}$.
From the full scattering amplitude, the analysing power is given by
\ba
  A_N\frac{d\sigma}{dt} =
         - 4 \pi [Im(\A_{nf})Re(\A_{sf})-Re(\A_{nf})Im(\A_{sf})],
\ea
each term having electromagnetic and hadronic contributions.

\begin{figure}
~\\~\\~\\
\epsfysize=3.cm
{\epsfbox{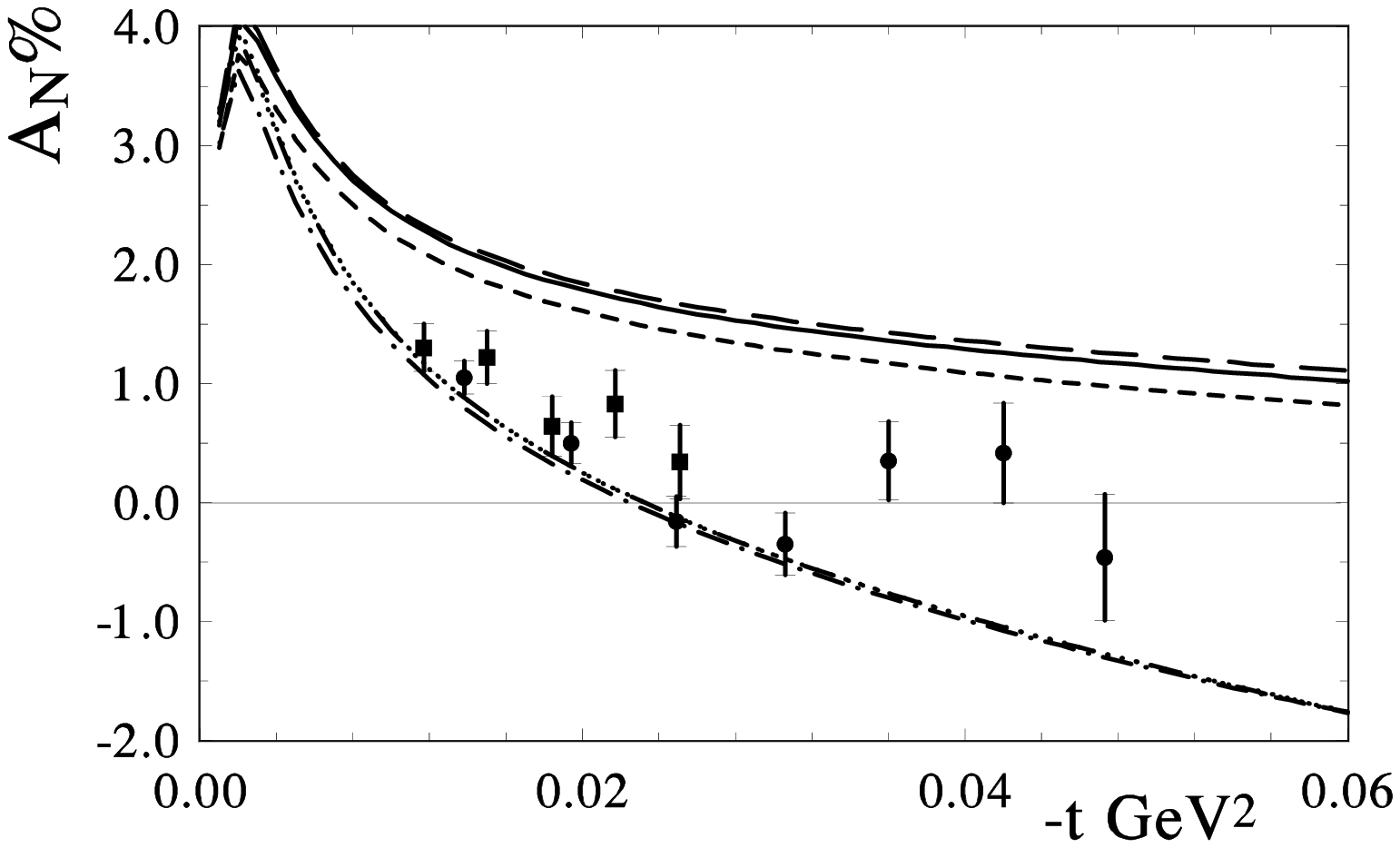}}
\epsfysize=3.cm
{\epsfbox{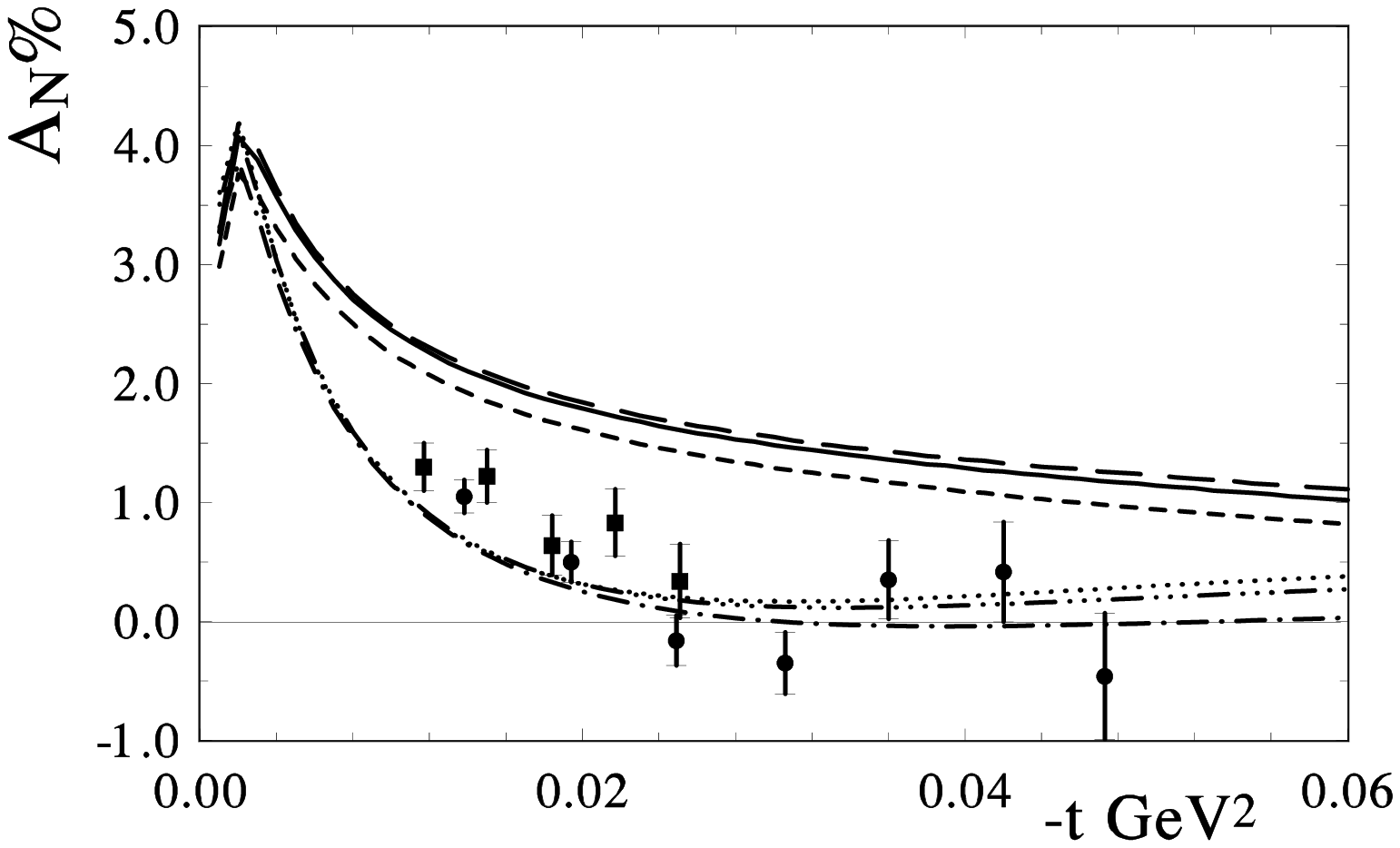}}
\caption{
Left:  $A_N$ with hadron-spin-flip amplitude in
case I ($B^{-} = B^{+}$) for $p_L = 24, \ 100, \   250 \  $GeV/$c$.
(dash-dot, dashed-dots and dots correspondingly). \ \
Right: same as left, but in
case II ($B^{-} =2 B^{+}$)  }
\end{figure}

In Fig. 2, we show  $A_N$ calculated for the two possible
slopes of the hadron spin-flip amplitude (cases I and II).
We see that in both cases we obtain a small energy dependence.
In case I, when $B^{-} = B^{+}$, $A_N$ decreases less with $|t|$ immediately
after the maximum. But at large $|t| \geq 0.01 $GeV$^2$ the behaviour of
$A_N$ is very different: we can obtain a zero for $A_N$ at
$|t| \approx 0.02$GeV$^2$, after which $A_N$ becomes negative and grows
in magnitude.
In case II,   when $B^{-} = 2 B^{+}$,  $A_N$ approaches zero, may become
slightly negative and then grows  positive again.

\begin{figure}
\epsfysize=3.cm
\centerline{\epsfbox{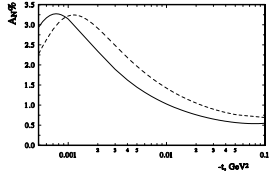}}
\caption{ $A_N$ for $pp$-scattering
(without hadron-spin flip amplitude) at
$\sqrt{s}=14 \ $TeV
(hard and dashed lines - with and without saturation)    }
\end{figure}

When we come to super-high energies, there may be
some additional effects connected with the
saturation of the unitarity bound for some values of the impact parameter.
This can lead
to different values for the analysing power. We calculate such effects
in the framework of a model with two simple poles, the second one
corresponding to
contribution of the hard pomeron.  Fig. 4 shows that the
effect of the saturation moves
the maximum of $A_N$ to smaller values of $t$.
Note that if we observe $A_N$ only in a region of $t$ after
the maximum, it will seem smaller, and we may wrongly conclude
that such effect comes from the hadron spin-flip amplitude.
This situation can also occur in $pA$-scattering at lower energies.

In conclusion, accurate measurements of the analysing power in
the  Coulomb-hadron interference region will reveal
the structure of the hadron spin-flip amplitude. This in turn
will give us further information about the
behaviour of the hadron interaction potential at large distances.
A definite example of the interplay between long and short distances
can be found in the peripheral dynamic model \ci{zpc,yaf-str},
which takes into account the contribution of the hadron
interaction at large  distances, and in which
the calculated hadron spin-flip amplitude
leads to a difference in the slopes of the ``reduced'' spin-flip and
spin-non-flip amplitudes at small momentum transfer.

\end{document}